\documentstyle[12pt]{article}
\input epsf.sty
\newcommand{\beqn}{\begin{equation}}
\newcommand{\eeqn}{\end{equation}}

\newcommand{\tht}[1]{\theta_#1}
\newcommand{\thdt}[1]{\dot{\theta_#1}}
\newcommand{\ttd}{\theta,\dot{\theta}}
\newcommand{\ld}[1]{\mbox {$\lambda_#1$}}
\newcommand{\vspc}{\vspace{.25in}}
\newcommand{\e}[1]{{{\bf \hat{e}}_#1}}

\begin{document}

\title{\bf
A comparative study of computation of Lyapunov spectra with different
algorithms
}

\author{
{\Large
\bf
K.Ramasubramanian {\normalsize and } M.S.Sriram
}\\
Department of Theoretical Physics, University of Madras,\\
Guindy Campus, Chennai 600 025, {\bf India}\\
E-mail: tphysmu@imsc.ernet.in}
\maketitle

\baselineskip 0.9truecm

\begin{abstract} In this paper we make a detailed numerical comparison
between three algorithms for the computation of the full Lyapunov spectrum
as well as the associated eigenvectors of general dynamical systems. They
are : (a) the standard method, (b) a differential formulation of the
standard method, and (c) a new algorithm which does not require rescaling
and reorthogonalization. We also bring out the relations among these
methods. Moreover, we give a simplified formulation of the new algorithm
when the dimensionality of the system is 4. We find that there is reasonable
agreement among the Lyapunov spectra obtained using the three algorithms in
most cases. However the standard method seems to be the most efficient
followed by the new method and the differential version of the standard
method (in that order), as far as the CPU time for the computation of the
Lyapunov spectra is concerned. The new method is hardly suitable for finding
the eigenvectors, whereas the other procedures give nearly identical
numerical results. \end{abstract}


\section{Introduction}

Extreme sensitivity to initial conditions is the commonly accepted defining
property of chaos in nonlinear systems. Lyapunov exponents which determine
the exponential rates at which nearby trajectories diverge on an average,
are the quantitative characteristics of a chaotic orbit. A dynamical system
of dimension $n$ has $n$ Lyapunov exponents and $n$ principal directions or
eigenvectors, corresponding to a set of nearby trajectories [1]. One of the
standard and popular methods to compute the Lyapunov spectrum of a dynamical
system involves a Gram-Schmidt Reorthonormalizaton (GSR) of the 'tangent
vectors' [2]. A differential version of this method has been formulated
which corresponds to a continuous GSR of the tangent vectors [3]. A
modification of this method with the introduction of a stability parameter
makes it dynamically stable, applicable to systems with degenerate spectra,
and reliable for computations [4]. Recently, a new algorithm for the
computation of Lyapunov exponents has been proposed, which has been claimed
to be valid even for evaluating partial Lyapunov spectra [5]. This is based
on the 'QR' method for the decomposition of the tangent map, (where Q is an
orthogonal matrix and R is an upper triangular matrix) which has been
studied by several authors [6]. It utilizes representations of orthogonal
matrices applied to the tangent map, and does not require the GSR procedure.
It has also been claimed that it has several advantages over the existing
methods, as it involves a minimum number of equations. In this paper we have
made a detailed comparison of the three algorithms as regards accuracy and
efficiency, by computing the full Lyapunov spectra of some typical nonlinear
systems with 2,3 and 4 variables. We also compare the performance of the
standard method with its differential version, in computing the Lyapunov
eigenvectors.

\vspc

In section 2, we outline the three methods with necessary details. We bring
out the relation between the differential version of the standard method and
the new procedure, by deriving the differential equations of the latter from
those of the former. It is difficult to use the new method with a standard
representation of orthogonal matrices when the number of dimensions of the
system is greater than 3. In section 3, we give a convenient representation
for them for $n=4$, by making use of the well-known fact that $SO(4) \sim
SO(3) \times SO(3)$ [7]. This simplifies the calculations considerably. In
section 4, we make a comparative study of the three algorithms for the
computation of Lyapunov spectra by taking up some typical 2, 3 and 4
dimensional systems. We have considered both dissipative and Hamiltonian
systems of some physical interest, for comparison. In section 5, we compare
the computation of the Lyapunov eigenvectors (which are local properties),
using these algorithms. In section 6, we make a few concluding remarks.
 
\section{Computation of Lyapunov exponents}
Consider an n-dimensional continuous-time dynamical system:
\beqn
{d{\rm \bf Z}\over dt} = {\rm \bf F}({\rm \bf Z},t),
\eeqn
where {\bf Z} and {\bf F} are n-dimensional vector fields. To determine the
n Lyapunov exponents of the system, corresponding to some initial condition
${\rm \bf Z(0)}$, we have to find the long term evolution of the axes of an
infinitesimal sphere of states around {\rm \bf Z(0)}. For this, consider the
tangent map given by the set of equations,

\beqn
{d\delta {\rm \bf Z}\over dt} = {\rm \bf J.\delta Z},
\eeqn

where {\rm \bf J} is the $ n \times n $ Jacobian matrix with 

\beqn
 J_{ij} = {\partial F_{i}\over \partial Z_{j}}. 
\eeqn

\vspace{.25in}

A solution of equation (2) can be formally written as 

\beqn 
\delta {\rm \bf Z({\rm t})} =  {\bf M}({\bf Z}(t),t) \delta {\rm \bf Z({\rm
0}}), 
\eeqn

where {\rm \bf M(Z({\rm t}), {\rm t})} is the tangent map whose
evolution equation is easily seen to be

\beqn
{d{\rm \bf M}\over dt} = {\rm \bf J.M}.
\eeqn

In the following, we give a brief description of the procedures for
computing the $ n $ Lyapunov exponents of the system using $(a)$ the
standard method, $(b)$ the differential version of the standard method and
$(c)$ the new method based on the 'QR' decomposition of $ M $, which
dispenses with the tangent vectors $\delta {\rm \bf Z}$, and in a sense,
computes the exponents directly.

\vspc

{\large \bf (a) The Standard method}

Let $ \ld{1}, \ld{2}, \ldots, \ld{n}$ be the $n$ Lyapunov exponents of the
system in a decreasing sequence, $ \ld{1} \geq \ld{2} \ldots \geq \ld{n}$.
In the standard method [2] one first chooses $n$ orthogonal tangent vectors
as initial conditions for eq.(2). The standard choice is $ \e{1}(0) =
(1,0,0, \ldots) $; $ \e{2}(0) = (0,1,0,0, \ldots) $, etc.. Eq.(2) is
then solved upto time $\tau$ for each of the initial conditions yielding
vectors $\bf v_{1}(\tau), v_{2}(\tau), \ldots v_{n}(\tau)$. These vectors
are orthonormalized using a Gram-Schimdt Reorthonormalization (GSR)
procedure to yield:

\begin{eqnarray}
\e{1}(\tau) & = & {\bf v_{1}\over \|\bf v_{1}\|}, \nonumber \\
\e{2}(\tau) & = & {\bf v_{2} - (v_{2},\e{1}(\tau))\e{1}(\tau) \over 
		\| \bf v_{2} - (v_{2},\e{1}(\tau))\e{1}(\tau) \|},
\end{eqnarray}
 
and so on. The norms in the denominators, denoted by $ N_{1}(1), N_{2}(1),
\ldots N_{n}(1),$ are stored for the computation of Lyapunov exponents. The
procedure is repeated for a subsequent time $ \tau $ of integration using $
\e{i}(\tau) $ as initial conditions for eqn.(2).  The resulting vectors $ \bf
v_{i}(2 \tau)$, are again orthonormalized using a GSR procedure to yield
orthonormal tangent vectors $ \e{i}(2 \tau), \; i=1, \ldots ,n $ and the
norms $ N_{1}(2), N_{2}(2), \ldots N_{n}(2)$. After $ r $ iterations, we get
the orthonomal set of vectors $ \e{i}(r \tau), i=1, \ldots ,n $ at time $ t =
r\tau$. The Lyapunov exponents are

\beqn
\ld{i} = \lim_{r \rightarrow \infty} {\sum_{m=1}^{r} \log N_{i}(m) 
		\over r \tau} 
\eeqn

This is due to the following reason. Since GSR never affects the direction
of the first vector in a system, this vector tends to seek out the direction
in the tangent space, which is most rapidly growing and its norm is
proportional to $ e^{\lambda_{1}t} $ for large $ t $. The second vector has its
component along the direction of the first vector removed and its norm would
be proportional to $ e^{\lambda_{2}t} $ for large $ t $ and so on.

\vspc

It is to be noted that we have to integrate $ n(n+1) $ coupled equations in
this method, as there are $ n $ equations for the fiducial trajectory in (1)
and $ n $ copies of the tangent map equations in (2).

\vspc

{\large \bf (b) Differential version of the standard method}

\vspc

In this method [3], the orthonormal set of vectors $ \e{i}(t) $ are obtained
by solving differential equations set up for them, instead of resorting to
the GSR at discrete steps. Rather, GSR is incorporated in the procedure
itself. It can be shown that

\beqn
{d \over dt}\e{i}(t) = {\bf G}\e{i} - G_{ii}\e{i} - 
			\sum_{j=1}^{i-1}(G_{ij} + G_{ji}) \e{j},
\eeqn

where {\bf G = J} is the Jacobian matrix introduced in eq.(2) and 

\beqn
G_{ij} = (\e{i}(t),{\bf J}({\bf Z}(t))\;\e{j}(t)),
\eeqn

that is, $G_{ij}$ are the matrix elements of the Jacobian in the basis
$\e{i}(t)$. Now let $\e{i}(0)$ evolve to $ {\bf e}_{i}(t)$.

\beqn
{\bf e}_{i}(t) = {\bf M}({\bf Z}(t),t) \;\e{i}(0),
\eeqn

In fact, $\e{i}(t)$ is the orthonormalized set corresponding to ${\bf
e}_{i}(t) \; i= 1, \ldots n $. Define

\beqn
d_{ij} = ({\bf e}_{i}(t), \e{j}(t)).
\eeqn

The GSR procedure ensures that $d_{ij}$ is a lower triangular matrix:

\beqn
d_{ij} = 0, \;\; i<j.
\eeqn

It can be shown that 

\beqn
\dot{d}_{ii} = G_{ii}d_{ii}, \;\; i=1,\ldots,n,
\eeqn

and that,

\beqn
d_{ii} \approx e^{\ld{i}t}
\eeqn

for large t. That is,

\beqn
\ld{i} = \lim_{t \rightarrow \infty} {1\over t} \log d_{ii}. 
\eeqn

The Lyapunov exponents are computed by solving the coupled equations (1),
(8) and  (13) in this method. As there are $n^{2}$ equations for the $n$
components each of the orthonormal vectors $ \e{i}(t) $ in (8), $ n $
equations for $ d_{ii} $ in (11), apart from the $ n $ equations for the
fiducial trajectory in (1), we have to integrate $ n(n+2) $ coupled
equations in this method.

\vspc

In practice, this procedure is not numerically 'stable', as the set $
\e{i}(t) $ may not remain orthonormal under the time evolution. In
particular $ \Delta_{ij}=(\e{i}(t),\e{j}(t)) - \delta_{ij},\; 1 \leq i,j
\leq n $ may not all vanish. Moreover, the method is not applicable to
systems with degenerate exponents. These are remedied by a modification of
the method, using a stability parameter $ \beta $ [4]. We replace $G_{ii}$
by $G_{ii} + \beta((\e{i},\e{i}) - 1)$ and $G_{ij}$ by $ G_{ij} + \beta
(\e{i},\e{j}), i \neq j $ in equations (8) and (13). Though it has been
shown that the method is strongly stable when $\beta > - \ld{n} $, where $
\ld{n} $ is the lowest exponent, it is found in certain problems, that $\beta
$ has to be significantly larger than $ -\ld{n} $ in practice. Moreover, it
may be pointed out that this method requires prior knowledge of the lowest
Lyapunov exponent $ \ld{n} $ for the computation of the complete spectrum $
\ld{i}$. If an arbitrarily high value is assigned to $\beta$, one ends up
with an arithmetic overflow problem during computations.

\vspc

{\large \bf (c) The New method based on a 'QR' decomposition of M}

\vspc

The new algorithm [5] is based on a 'QR' decomposition of M, where Q is an
orthogonal matrix and R is an upper triangular matrix. This results in a set
of coupled differential equations for the Lyapunov exponents along with the
various angles parametrising the orthogonal matrices. In this subsection we
derive these equations from the differential version of the standard method
considered in the previous subsection .

Consider the tangent map matrix ${\bf M}$. From eq.(10),

\beqn
M_{ij}= (\e{i}(0), {\bf M} \e{j}(0)) = (\e{i}(0), {\bf e}_{j}(t)) 
\eeqn

As $ \e{j}(t) $ form an orthonormal set of vectors, we have from eq.(11), 

\beqn
{\bf e}_{j}(t) = \sum  \e{k}(t) d_{jk}.
\eeqn

Hence,

\beqn
M_{ij}= \sum_{k} (\e{i}(0),  \e{k}(t)) d_{jk}.
\eeqn
 
Define the matrices $ Q $ and $ R $ by

\begin{eqnarray}
Q_{ij} & = &(\e{i}(0), \e{j}(t)) = (\e{j}(t))_{i},\\
{\rm and} && \nonumber \\
R_{ij} & = & d_{ji}.
\end{eqnarray}

Hence,

\beqn
M = QR.
\eeqn

Clearly the columns of $Q$ are the orthonormal vectors $ \e{j}(t) $, and $Q$
is an orthogonal matrix. As $ d $ is a lower triangular matrix, $ R $ is an
upper triangular matrix. 

\vspc

Now $ G_{ij} $ and $ J_{ij} $ are the matrix elements of the Jacobian in the
orthonormal bases $ \e{i}(t) $ and $ \e{i}(0) $ respectively and related by
a rotation transformation represented by $Q$.

\vspc

Introducing complete sets of states at the appropriate places, we have

\begin{eqnarray}
G_{ij} & = &(\e{i}(t),{\bf J} \e{j}(t)) \nonumber \\
       & = & \sum_{k,l} (\e{i}(t),\e{k}(0))\;(\e{k}(0),{\bf J}\e{l}(0))
		(\e{l}(0),\e{j}(t)) \nonumber \\
       & = & \sum_{k,l} \tilde{Q}_{ik}J_{kl}Q_{lj} = (\tilde{Q}JQ)_{ij}.
\end{eqnarray}

Taking the scalar product of eq.(8) with $\e{j}(0)$ and making appropriate
changes of indices, we have

\begin{eqnarray}
{d \over dt}Q_{jk} & = & {d \over dt}(\e{j}(0),\e{k}(t)) \nonumber \\
		   & = & (\e{j}(0),{\bf J} \e{k}(t)) - G_{kk}(\e{j}(0),\e{k}(t))
		         - \sum_{l=1}^{k-1} (G_{kl}+G_{lk})
			   (\e{j}(0),\e{l}(t)) \nonumber \\
 		   & = & (\e{j}(0),{\bf J} \e{k}(t)) - G_{kk}Q_{jk}
		         - \sum_{l=1}^{k-1} (G_{kl}+G_{lk})Q_{jl}.
\end{eqnarray}

As all the quantities are real,

\beqn
\tilde{Q}_{ij} = Q_{ji} = (\e{j}(0),\e{i}(t)) = (\e{i}(t),\e{j}(0)).
\eeqn

Multiplying eq.(23) by $Q_{ij}$ on the right and using the fact that

\begin{eqnarray}
\tilde{Q}_{ij}(\e{j}(0),{\bf J} \e{k}(t)) & = & \sum_{j}(\e{i}(t),\e{j}(0))
			   (\e{j}(0),{\bf J} \e{k}(t)) \nonumber \\
 		   & = & (\e{i}(t),{\bf J} \e{k}(t)) = G_{ik}
\end{eqnarray}

we find

\begin{eqnarray}
(\tilde{Q}{d \over dt}Q)_{ik} & = & \tilde{Q}_{ij}{d \over dt}Q_{jk} \nonumber \\
		   & = & G_{ik} - G_{kk}\sum_{j}\tilde{Q}_{ij}Q_{jk}
		         - \sum_{j}\sum_{l=1}^{k-1} (G_{kl}+G_{lk})
			   \tilde{Q}_{ij}Q_{jl} \nonumber \\
		   & = & G_{ik} - G_{kk}\delta_{ik}
		         - \sum_{l=1}^{k-1} (G_{kl}+G_{lk}) \delta_{il},
\end{eqnarray}

as Q is an orthonormal matrix.

\vspc

Again, $\tilde{Q}{d \over dt}Q$ is an antisymmetric matrix as $Q$ is orthogonal
and it is sufficient to consider $ i > k $. In this case, the last term
vanishes and we obtain,

\beqn
(\tilde{Q}{d \over dt}Q)_{ik} = G_{ik} = (\tilde{Q}JQ)_{ik}, \; i>k.
\eeqn

Q being an orthogonal matrix is characterised by $ n(n-1) \over 2 $ angles
and we obtain differential equations for these angles. From eqs.(13)
and (14), the differential equations for the Lyapunov exponents are

\beqn
{d \over dt}(\ld{i}t) = G_{ii} = (\tilde{Q}JQ)_{ii}. 
\eeqn

In this method, we have essentially traded the orthonormal vectors $\e{i}(t)$
for the orthogonal matrix $ Q $ parametrized by the $ n(n-1) \over 2 $
angles. We have to solve the coupled eqs.(1), (27) and (28) in this
procedure to obtain the Lyapunov exponents. We have to integrate $ n+{n(n-1)
\over 2} + n = {n(n+3) \over 2}$ coupled equations in this method.

\section{A convenient representation for {\rm Q} and simplification of $
\tilde{Q}\dot{Q}$ for $n=4$}

In [5], the explicit representation of the orthogonal matrix $Q$ used is
the one in which it is represented as a product of $n(n-1) \over 2$
orthogonal matrices, each of which corresponds to a simple rotation in the
$i\/-\/j^{th}$ plane $(i<j)$. Thus $Q$ 

\beqn
\begin{array}{l}
Q = O^{(12)}O^{(13)}O^{(14)} \ldots O^{(1n)}O^{(23)} \ldots
        O^{(n-2,n-1)}O^{(n-1,n)} \\ 
{\rm where}\\
\begin{array}{rlcl}
O_{kl}^{(ij)}& = 1 & {\rm if} & k=l \neq i,j;\\
	     & = \cos\tht{{ij}} & {\rm if} & k=l=i{\rm or}j;\\
	     & = \sin\tht{{ij}} & {\rm if} & k=i,l=j;\\
	     & = -\sin\tht{{ij}} & {\rm if} & k=j,l=i;\\
	     & = 0 &  & {\rm otherwise}.	
\end{array}
\end{array}
\eeqn
In terms of the group generators, $O^{(ij)}$ can be written as 
\beqn
O^{(ij)} = e^{\tht{{ij}}(t_{ij})},
\eeqn

where the generator $t_{ij}$ is represented by

\beqn
(t_{ij})_{kl} = \delta_{ik}\delta_{jl} - \delta_{il}\delta_{jk},
\eeqn 

The generators satisfy the commutation relations,

\beqn
[t_{ij},t_{mn}] = \delta_{in}t_{jm} + \delta_{jm}t_{in} -
		 	\delta_{im}t_{jn} - \delta_{jn}t_{im}.
\eeqn 

The above representation for $Q$ is conceptually simple and works very well
for $n=2$ and 3 [5]. However, for $n>3$, it is hardly suitable for practical
computations of Lyapunov exponents. This is because the expressions for
$\tilde{Q}\dot{Q}$ and $\tilde{Q}{\bf J} Q$ are very lengthy and
unmanageable even for $n=4$.

In the present work, we employ a representation for Q, which simplifies the
calculations and numerical computations for $n=4$. This is based on the
well-known fact that $SO(4)\sim SO(3) \times SO(3)$ [7]. From the generators
$t_{ij}$ we construct the following combinations:

\beqn
\begin{array}{cc}
M_{1} = {1\over2}(t_{23}+t_{14}),  &  N_{1} = {1\over2}(t_{23}-t_{14})\\\\
M_{2} = {1\over2}(t_{31}+t_{24}),  &  N_{2} = {1\over2}(t_{31}-t_{24})\\\\
M_{3} = {1\over2}(t_{12}+t_{34}),  &  N_{3} = {1\over2}(t_{12}-t_{34}).
\end{array}
\eeqn

Then it is easily verified that $M_{i}$ and $N_{i}$ generate two mutually
commuting $SO(3)$ algebras:

\beqn
\begin{array}{ccc}
[M_{i},M_{j}] = -\epsilon_{ijk} M_{k},
		& [N_{i},N_{j}] = -\epsilon_{ijk} N_{k}, & [M_{i},N_{j}]=0.
\end{array}
\eeqn

We write $Q$ as

\beqn
Q  = Q_{II}Q_{I},
\eeqn
where
\beqn
 Q_{II} = O^{(6)}O^{(5)}O^{(4)} = e^{\tht{6}N_{3}}
			e^{\tht{5}N_{2}}e^{\tht{4}N_{1}}
\eeqn
 and
\beqn
 Q_{I} = O^{(3)}O^{(2)}O^{(1)} = e^{\tht{3}M_{3}}
			e^{\tht{2}N_{2}}e^{\tht{1}M_{1}}.
\eeqn
Using
\beqn
	e^{X}Ye^{-X} = Y + [X,Y] + {1 \over 2!}[X,[X,Y]] + \ldots,
\eeqn
for any matrices $X$ and $Y$ and the commutation relations in equation (32),
it can be easily verified that

\beqn
\begin{array}{rl}
\tilde{Q}\dot{Q}& = \tilde{Q}_{I}\dot{Q_{I}}+\tilde{Q}_{II}\dot{Q_{II}}\\ \\

		& = [\thdt{1}+\thdt{3}\sin\tht{2}]M_{1} + 
		    [\thdt{2}\cos\tht{1}+\thdt{3}\sin\tht{1}\cos\tht{2}]M_{2}\\
		& + [\thdt{2}\sin\tht{2}+\thdt{3}\cos\tht{1}\cos\tht{2}]M_{3}+ 
		    [\thdt{4}+\thdt{6}\sin\tht{5}]N_{1} + \\
		& + [\thdt{5}\cos\tht{4}-\thdt{6}\sin\tht{4}\cos\tht{5}]N_{2}+ 
		[\thdt{5}\sin\tht{5}+\thdt{6}\cos\tht{4}\cos\tht{5}]N_{3}. 
\end{array}
\eeqn 
 
The explicit form of the matrices $M_{i} {\rm and} N_{i}$ can be found using
equations (31) and (33) and are written in terms of $2 \times 2$ blocks as
given below:

\beqn
\begin{array}{lll}
M_{1} = {1 \over 2}\left[\begin{array}{ccc}
	  0  & \vdots & \sigma_{1}   \\
	  \cdots &   \cdots & \cdots \\
	   -\sigma_{1} & \vdots & 0
	\end{array} \right], 	
&
M_{2} = {1 \over 2}\left[\begin{array}{ccc}
	  0  & \vdots & -\sigma_{3}   \\
	  \cdots &   \cdots & \cdots \\
	   \sigma_{3} & \vdots & 0
	\end{array} \right], 	
&
M_{3} = {1 \over 2}\left[\begin{array}{ccc}
	   i\sigma_{2} & \vdots & 0   \\
	  \cdots &   \cdots & \cdots \\
	   0 & \vdots & i\sigma_{2}
	\end{array} \right]\\ \\ 	

N_{1} = {1 \over 2}\left[\begin{array}{ccc}
	  0  & \vdots & -i\sigma_{2}   \\
	  \cdots &   \cdots & \cdots \\
	   -i\sigma_{2} & \vdots & 0
	\end{array} \right], 	
&
N_{2} = {1 \over 2}\left[\begin{array}{ccc}
	  0  & \vdots & -I   \\
	  \cdots &   \cdots & \cdots \\
	   I & \vdots & 0
	\end{array} \right], 	
&
N_{3} = {1 \over 2}\left[\begin{array}{ccc}
	   i\sigma_{2} & \vdots & 0   \\
	  \cdots &   \cdots & \cdots \\
	   0 & \vdots & -i\sigma_{2}
	\end{array} \right]. 	
\end{array}
\eeqn
Here I is the $2\times2$ identity matrix and $\sigma_{1},\sigma_{2}$ and
$\sigma_{3}$ are the Pauli matrices:
\beqn
\begin{array}{cccc}
I = \left[\begin{array}{cc}
	   1 & 0 \\
	   0 & 1
	\end{array} \right], 	
&\sigma_{1} =\left[\begin{array}{cc}
	   0 & 1 \\
	   1 & 0
	\end{array} \right], 	
& \sigma_{2} =\left[\begin{array}{cc}
	   0 & -i \\
	   i & 0
	\end{array} \right], 	
& \sigma_{3} =\left[\begin{array}{cc}
	   1 & 0 \\
	   0 & -1
	\end{array} \right] .	
\end{array}
\eeqn

Then we find that 
\beqn
 \tilde{Q}\dot{Q} = \left( \begin{array}{cccc}
	  0          & -f_{1}(\ttd) & -f_{2}(\ttd)  & -f_{3}(\ttd)\\
	  f_{1}(\ttd) &  0	    & -f_{4}(\ttd)  & -f_{5}(\ttd)\\
	  f_{2}(\ttd) &  f_{4}(\ttd)  &  0          & -f_{6}(\ttd) \\
	  f_{3}(\ttd) &  f_{5}(\ttd)  & f_{6}(\ttd) & 0
	\end{array} \right),  	
\eeqn

where

\beqn
\begin{array}{rl}
	f_{1}& =-{1\over2}(\thdt{2}\sin\tht{1} + \thdt{3}\cos\tht{1}\cos\tht{2}+
		  \thdt{5}\sin\tht{4} +\thdt{6}\cos\tht{4}\cos\tht{5}),\\	
	f_{2}& ={1\over2}(\thdt{2}\cos\tht{1} - \thdt{3}\sin\tht{1}\cos\tht{2}+
		  \thdt{5}\cos\tht{4} - \thdt{6}\sin\tht{4}\cos\tht{5}),\\	
	f_{3}& =-{1\over2}(\thdt{1} + \thdt{3}\sin\tht{2}-
		     \thdt{4} - \thdt{6}\sin\tht{5}),\\	
	f_{4}& =-{1\over2}(\thdt{1} + \thdt{3}\sin\tht{2}+
		     \thdt{4} + \thdt{6}\sin\tht{5}),\\	
	f_{5}& ={1\over2}(-\thdt{2}\cos\tht{1}+\thdt{3}\sin\tht{1}\cos\tht{2}+
		  \thdt{5}\cos\tht{4} - \thdt{6}\sin\tht{4}\cos\tht{5}),\\	
	f_{6}&=-{1\over2}(\thdt{2}\sin\tht{1}+\thdt{3}\cos\tht{1}\cos\tht{2}-
		     \thdt{5}\sin\tht{4}-\thdt{6}\cos\tht{4}\cos\tht{5}).	
\end{array}
\eeqn 

Using equation(12), we find that the equations for $\dot{\tht{i}}$, split
neatly into two sets:

\beqn
\begin{array}{ccc}
 \left( \begin{array}{lll}
	  -1    & 0  & -\sin\tht{2}\\
	  0    & -\sin\tht{1} & -cos\tht{1}\cos\tht{2}\\	 
	  0    & \cos\tht{1} & -sin\tht{1}\cos\tht{2}	 
	\end{array} \right)  	
& \left(\begin{array}{c}
	   \thdt{1} \\ \thdt{2} \\ \thdt{3}
	\end{array} \right) =
& \left(\begin{array}{c}
	   G_{32} + G_{41} \\ G_{21} + G_{43} \\ G_{31} - G_{42}
	\end{array} \right) ,
\end{array} 	
\eeqn
and 
\beqn
\begin{array}{ccc}
 \left( \begin{array}{lll}
	  -1    & 0  & -\sin\tht{5}\\
	  0    & -\sin\tht{4} & -cos\tht{4}\cos\tht{5}\\	 
	  0    & \cos\tht{4} & -sin\tht{4}\cos\tht{5}	 
	\end{array} \right)  	
& \left(\begin{array}{c}
	   \thdt{4} \\ \thdt{5} \\ \thdt{6}
	\end{array} \right) =
& \left(\begin{array}{c}
	   G_{32} - G_{41} \\ G_{21} - G_{43} \\ G_{31} + G_{42}
	\end{array} \right) .
\end{array} 	
\eeqn

We also have
\beqn
{d \over dt}(\ld{i}t) = G_{ii}, \; i=1,\ldots,4,
\eeqn

from eq.(28). Hence, to find the Lyapunov exponents of a dynamical
system with 4 variables, we have to solve the evolution equations for the
system given by eq.(1) and the tangent map equations given by equations
(44), (45) and (46) after finding $ G \equiv \tilde{Q}\/ {\bf J}\/ Q$.

Any $4\times4$ matrix {\bf J} can be written as :

\beqn
J = \sum_{i=1}^{16}a_{i}X_{i},
\eeqn

where the 16 matrices $X_{i}$ are defined in terms of $2\times2$ blocks as

\beqn
\begin{array}{llll}
X_{1} = \left[\begin{array}{cc}
	   I & 0 \\
	   0 & I
	\end{array} \right], 	
&X_{2} =\left[\begin{array}{cc}
	   -I & 0 \\
	   0 & I
	\end{array} \right], 	
&X_{3} =\left[\begin{array}{cc}
	   \sigma_{3} & 0 \\
	   0 & \sigma_{3}
	\end{array} \right], 	
&X_{4} =\left[\begin{array}{cc}
	   -\sigma_{3} & 0 \\
	   0 & \sigma_{3}
	\end{array} \right]\\ \\
X_{5} =\left[\begin{array}{cc}
	   \sigma_{1} & 0 \\
	   0 & \sigma_{1}
	\end{array} \right], 	
&X_{6} =\left[\begin{array}{cc}
	   -\sigma_{1} & 0 \\
	   0 & \sigma_{1}
	\end{array} \right],
&X_{7} =\left[\begin{array}{cc}
	   i\sigma_{2} & 0 \\
	   0 & i\sigma_{2}
	\end{array} \right], 	
&X_{8} =\left[\begin{array}{cc}
	   i\sigma_{2} & 0 \\
	   0 & -i\sigma_{2}
	\end{array} \right]\\ \\
X_{9} = \left[\begin{array}{cc}
	   0 & I \\
	   I & 0
	\end{array} \right], 	
&X_{10} =\left[\begin{array}{cc}
	   0 & I \\
	  -I & 0 
	\end{array} \right], 	
&X_{11} =\left[\begin{array}{cc}
	  0 &  \sigma _{3}  \\
	   \sigma_{3} & 0
	\end{array} \right], 	
&X_{12} =\left[\begin{array}{cc}
	   0 & \sigma_{3}  \\
	   -\sigma_{3} & 0
	\end{array} \right]\\ \\
X_{13} =\left[\begin{array}{cc}
	  0 &  \sigma _{1}  \\
	   \sigma_{1} & 0
	\end{array} \right], 	
&X_{14} =\left[\begin{array}{cc}
	   0 & \sigma_{1}  \\
	   -\sigma_{1} & 0
	\end{array} \right],
&X_{15} =\left[\begin{array}{cc}
	   0 & -i\sigma_{2}  \\
	   -i\sigma_{2} & 0
	\end{array} \right], 	
&X_{16} =\left[\begin{array}{cc}
	   0 & -i\sigma_{2}  \\
	   i\sigma_{2} & 0
	\end{array} \right].\\ \\
\end{array}
\eeqn

It is easy to find commutators $[X_{i},M_{j}]$ and $[X_{i},N_{j}]$ from
eqs.(40) and (48). Then, using eqs. (35--38), we can obtain

\beqn
G = \tilde{Q}\/{\bf J}\/Q. 
\eeqn

\section{A comparative study of the three algorithms for the computation of
Lyapunov spectra}

The standard algorithm involves an explicit GSR for finding the orthonormal
set $ \e{i}(t)$ and the Lyapunov spectrum. The differential version
considered in section 3(b) amounts to computing the spectrum with continuous GSR.
Here explicit GSR is avoided as it is incorporated in the method. However,
the differential equations for $\e{i}(t)$ in this method are nonlinear, as
they involve $(\e{i}(t), {\bf J} \e{j}(t)) $ in the RHS, in contrast to the
standard method which uses the linearized equations for $\delta Z$ directly.
In the new method, one deals directly with the orthogonal matrix relating
$\e{i}(t)$ and $\e{i}(0)$. It uses a minimal number of variables and
rescaling and reorthogonalization are eliminated. However, in this method,
the evolution equations for the angles and Lyapunov exponents are highly
nonlinear involving sines and cosines of the angles. Hence it is not clear
`a priori' which method is `superior' and there is a need to compare the
efficiency and accuracy of the three methods.  That is the subject matter of
the present investigation. Here we consider some typical nonlinear systems
of physical interest with $n=2,3$ and $4$. The driven Van der Pol oscillator
is taken as an example of a two dimensional system, whereas the standard
Lorenz system is chosen for $n=3$. For $n=4$, we consider the coupled
quartic oscillators and anisotropic Kepler problem as examples of
conservative Hamiltonian systems and the R\"ossler hyperchaos system as an
example of a dissipative system. We give the differential equations for
these dynamical systems in the following.

\vspace{.25in}

1. Driven Van der Pol oscillator ($n=2$):

\beqn
{d \over dt} \left( \begin{array}{c} z_{1} \\ z_{2}
	   \end{array} \right) =
	\left( \begin{array}{c} z_{2} \\ 
		-(d(1-z_{1}^2) z_{2} - z_{1} + b \cos \omega t,
	   \end{array} \right) .
\eeqn 
where $b$ and $d$ are parameters and $\omega$ is the driven frequency. In
our numerical work we have chosen $d=-5.0, b=5.0$ and $\omega=2.47$ as the
parameter values.

\vspace{.25in}

2. Lorenz system ($n=3$):

\beqn
{d \over dt} \left( \begin{array}{c} z_{1} \\ z_{2}\\z_{3}
	   \end{array} \right) =
	\left( \begin{array}{c} \sigma (z_{2} - z_{1})\\ 
		z_{1}(\rho - z_{3}) - z_{2}\\
		z_{1} z_{2} - \beta z_{3}
	   \end{array} \right) .
\eeqn 
This system is too well-known to require any further discussion. For
computations we set $ \sigma = 10.0, \rho=28.0$ and $\beta={8\over3}$.

\vspace{.25in}

3. Coupled quartic oscillators ($n=4$):

This is a conservative system and the Hamiltonian is given by

\beqn
H={z_{3}^2\over2}+{z_{4}^2\over2}+z_{1}^4+z_{2}^4+\alpha z_{1}^2z_{2}^2,
\eeqn

where $z_{1}$ and $z_{2}$ are the canonical coordinates, $z_{3}$ and $z_{4}$
are the corresponding momenta and $\alpha$ is a parameter. The Hamiltonian
in eq.(53) finds applications in high energy physics [8], to mention just
one example.  The equations of motion are:

\beqn
{d \over dt} \left( \begin{array}{c} z_{1} \\ z_{2}\\z_{3}\\z_{4}
	   \end{array} \right) =
	\left( \begin{array}{c} z_{3} \\ z_{4}\\ 
		-(4z_{1}^3 + 2 \alpha z_{1} z_{2}^2)\\
		-(4z_{2}^3 + 2 \alpha z_{1}^2 z_{2})
	   \end{array} \right) .
\eeqn 

This system is known to be integrable for $\alpha = 0,2$ and 6 [9].

\vspace{.25in}

4. Anisotropic Kepler problem ($n=4$):

\vspace{.25in}

The Hamiltonian of this system is given by:

\beqn
H={p_{\rho}^2 \over 2}+ \gamma {p_{z}^2 \over 2}-{e^2 \over \sqrt{\rho^2+z^2}}
\eeqn

where $\gamma$ is a number. 

\vspace{.25in}

The Hamiltonian given above describes the motion of an electron in the Coloumb
field in an anisotropic crystal, where its effective mass along the x-y
plane and z-direction are different [10]. $\gamma = 1$ corresponds to the
isotropic case and is integrable. When $\gamma \neq 1$, the system is
non-integrable. Because of the singularity at $\rho=z=0$, the Hamiltonian in
the above form is hardly suitable for numerical integration. For this we
choose $z_{1}=\sqrt{\rho + z}$ and $z_{2}=\sqrt{\rho - z}$ as the canonical
variables. We can find the corresponding canonical momenta $z_{3}$ and
$z_{4}$ in terms of $p_{\rho}$ and $p_{z}$. We also use a re-parametrized
time variable $\tau$ defined by $dt = d\tau(z_{1}^2+z_{2}^2)$.

\vspace{.25in}

The original Hamiltonian with the old variables and energy $E$ corresponds
to the following Hamiltonian with $H' = 2$ in terms of the new variables
[11]:

\beqn
H'=2={1 \over 2}(z_{3}^2+z_{4}^2)-E(z_{1}^2+z_{2}^2)+
	(\gamma-1){(z_{1}z_{3}-z_{2}z_{4})^2 \over 2(z_{1}^2+z_{2}^2)}.
\eeqn

The equations of motion resulting from this are:

\beqn
{d \over dt} \left( \begin{array}{c} z_{1} \\ z_{2}\\z_{3}\\z_{4}
	   \end{array} \right) =
	\left( \begin{array}{c} z_{3} + 
	(\gamma-1)z_{1}{(z_{1}z_{3}-z_{2}z_{4}) \over (z_{1}^2+z_{2}^2)}\\ 
   z_{4}-(\gamma-1)z_{2}{(z_{1}z_{3}-z_{2}z_{4}) \over (z_{1}^2+z_{2}^2)}\\ 
  2Ez_{1}-(\gamma-1){(z_{3}^2z_{1}z_{2}^2+z_{2}z_{3}z_{4}(z_{1}^2-z_{2}^2)
	 	-z_{2}^2z_{1}z_{4}^2)\over (z_{1}^2+z_{2}^2)}\\ 
  2Ez_{2}-(\gamma-1){(z_{4}^2z_{2}z_{1}^2-z_{1}z_{3}z_{4}(z_{1}^2-z_{2}^2)
	 	-z_{1}^2z_{2}z_{3}^2)\over (z_{1}^2+z_{2}^2)}
	   \end{array} \right) .
\eeqn 

We have chosen $\gamma=0.61$ for computational purposes.

\vspace{.25in}

5. R\"ossler hyperchaos system (n=4):

\vspace{.25in}

This is a dissipative system and an extension of the three dimensional
R\"ossler attractor [9]. It is described by the equations:

\beqn
{d \over dt} \left( \begin{array}{c} z_{1} \\ z_{2}\\z_{3}\\z_{4}
	   \end{array} \right) =
	\left( \begin{array}{c} -(z_{2}+z_{3}) \\ z_{1}+az_{2}+z_{4}\\ 
		b + z_{1} z_{3}\\ cz_{4} - d z_{3}
	   \end{array} \right) ,
\eeqn 
where $a,b,c$ and $d$ are parameters whose values are taken to be 0.25, 3.0,
0.05 and 0.5 respectively for our computations.

\vspace{.25in}

In all these cases, the full Lyapunov spectrum is computed using the three
methods. The time of integration is chosen to ensure reasonable convergence
of the Lyapunov exponents. In most of the cases the time of integration was
$t=1,00,000$ (the exceptions are the anisotropic Kepler problem and the 
R\"ossler hyperchaos system using the differential version of the standard
method due to the problem of numerical overflow). For all the systems, we
have used a variable step-size Runge Kutta routine (RKQC) for integration,
with an error tolerance, $\epsilon \sim 10^{-6}-10^{-8}$. All the
computations were performed on a DEC Alpha based workstation running
OpenVMS. The CPU time taken for each system with each of the algorithms was
noted. This is the actual time taken by the CPU to accomplish a specific
process (independent of the other processes running in the system). The
details of the comparison between the two methods are summarized in table 1.

\vspc

It may be noticed that all the  methods yield essentially the same Lyapunov
spectrum. For any autonomous dynamical system, one of the Lyapunov exponents
has to be zero (corresponding to the difference vector $\delta z$ lying
along the trajectory itself). For the Lorenz system, the R\"ossler
hyperchaos system (both dissipative) and the coupled quartic oscillators,
this condition is satisfied by all the algorithms. For the anisotropic
Kepler problem all the methods fail the test. This aspect needs to be
studied further. For Hamiltonian systems, for every eigenvalue $\lambda$,
there is an eigenvalue $-\lambda$. This symmetry is respected by all the
algorithms. For the coupled quartic oscillators, all the exponents should be
zero corresponding to the integrable case of $\alpha=6$. This is indeed
satisfied by all the algorithms. In Fig.1 we give plots of Lyapunov
exponents as functions of time, for a typical case. Again, there is no
significant difference between the three algorithms as far as the
convergence of the Lyapunov exponents is concerned. It is noteworthy that
the differential method works well for even systems with degenerate spectra
like the coupled quartic oscillators.

\vspc

On the whole, the standard method seems to have an edge over the new method
as far as the CPU time for the computation of the Lyapunov sepctrum is
concerned. The differential version of the standard method generally
consumes more CPU time compared to the other two methods. For some systems
like the anisotropic Kepler problem and the R\"ossler hyperchaos system,
there are numerical overflow problems, whatever be the values of $\beta$ and
the error tolerance $\epsilon$ one chooses for this algorithm. In fact, it
appears that the value of $\beta$ has to be significantly higher than
$-\ld{n}$ (indicated by the stability analysis) for these systems, for
reasonable convergence.

\vspc

For the system of coupled quartic oscillators, the CPU time is
abnormally high for the new method, corresponding to the nonintegrable case
of $\alpha=8$. This is true both for small and large energies. For large
energies ($\sim$ 25000), since the energy varied by $\sim 15$ when we used
the RKQC routine, we also used a symplectic procedure which eliminates
secular variations in the energy [13]. With this routine, the CPU times were
nearly the same for both the methods. However the new method yields poor
results for the Lyapunov spectrum. For instance corresponding to the initial
condition $z_{1}=7.0$, $z_{2}=7.0$, $z_{3}=5.0$ and $z_{4}=4.0$, the
Lyapunov spectrum computed using the new and the standard methods are
(1.5506, 0.3254, -0.3261, -1.5499) and (1.5205, 0.0001, -0.0001, -1.5205)
respectively. The differential version of the standard method led to a
numerical overflow problem, corresponding to this initial condition.

\vspc

In the standard method, after solving for the fiducial trajectory, the
equations for the tangent flow are linearized equations. In method (b),
corresponding to continuous GSR, these equations are nonlinear. In the new
method, these equations are replaced by the equations for the angles
determining the principal axes or the bases associated with the Lyapunov
spectrum and the Lyapunov exponents. These equations involving sines and
cosines of the angles are highly nonlinear. For dissipative systems this
nonlinearity does not pose a problem. However in many cases, this
nonlinearity renders the differential version of the standard method and the
new method less efficient and can even lead to inaccuracies, in strongly
chaotic situations.
 
\section{Lyapunov eigenvectors}

Earlier we had defined the matrix $d_{ij}$ as:

\beqn
d_{ij} = ({\bf e}_{i}(t), \e{j}(t)).
\eeqn

Consider the quantities

\begin{eqnarray}
\bar{d}_{ij} & = & {d_{ij} \over d_{jj}}, \; i \geq j, \nonumber \\
		& = & 0,		i < j.
\end{eqnarray}

Define the vectors $\bar{d}_{i}$ as

\begin{eqnarray}
\bar{d}_{1} 	& = & (\bar{d}_{11}, \bar{d}_{21},\bar{d}_{31},\ldots), \nonumber \\
\bar{d}_{2} 	& = & (0, \bar{d}_{22},\bar{d}_{32}, \ldots), etc.
\end{eqnarray}

Let ${\bf D_{1},D_{2}, \ldots ,D_{n},}$ be the orthonormal set of vectors
obtained form ${\bar d}_{i}'s$ by the Gram-Schmidt procedure, starting with
$\bar{d}_{1}$. It can now be shown that ${\bf D_{1},D_{2}, \ldots ,D_{n},}$
are the eigenvectors of $\tilde{M} M$ or the Lyapunov eigenvectors
corresponding to the eigenvalues $\ld{1}, \ld{2}, \ldots, \ld{n}$ [3]. In
this section, we consider the compuation and convergence of these
eigenvectors corresponding to the systems considered in section 4.

\vspc

In the standard method, we have to compute ${\bf e}_{i}(t)$ and $\e{i}(t)$
separately to obtain $\bar{d}_{ij}$. As all the vectors ${\bf e}_{j}(t)$
tend to align along $\e{1}$, both $d_{ij} = {\bf e}_{i}(t).\e{j}(t)$ and
$d_{jj} = {\bf e}_{j}(t).\e{j}(t)$ would tend to zero for $j > 1$. As
$\bar{d}_{ij}$ is the ratio of $d_{ij}$ amd $d_{jj}$, it would be difficult
to compute them for large $t$, in this method. Even then, the procedure
seems to give reasonable results for all the systems, we have considered. 

\vspc

In the differential version of the standard method, it has been shown that
$\bar{d}_{ij}$ satisfy the following differential equations [3]:

\beqn
{d \over dt}\bar{d}_{ij} = \sum_{k= j+1}^{i}{d_{kk} \over d_{jj}}
				\bar{d}_{ik}(G_{jk}+ G_{kj}), \; i>j.
\eeqn

So the eigenvectors are obtained by direct integration of these equations.
This procedure does not pose any problem as we do not come across division
by small numbers here. Indeed we find that the eigenvectors converge much
more rapidly than the Lyapunov exponents in all the cases, as anticipated by
Goldhirsch, et al.[3].

\vspc 

In the new method, the orthonormal vectors $\e{i}(t)$ are just the columns
of the orthogonal matrix $Q$. However, it is not straightforward to compute
${\bf{e}}_{i}(t)$ in this method. So we do not consider this method further
here. 

\vspc

We summarise the results for the Lyapunov eigenvectors in table 2 for the
same systems with the same parameters and initial conditions as in table 1.
As remarked earlier, the vectors converge sufficiently fast and the two
methods yield essentially identical results. Now for a Hamiltonian system,
the tangent map matrix $M$ satisfies the 'sympletic condition':

\begin{eqnarray}
\tilde{M}SM & = &S,\\ 
{\rm with} \nonumber \\
 S & = & \left[\begin{array}{ccc}
	  0  & \vdots & I   \\
	  \cdots &   \cdots & \cdots \\
	  -I & \vdots & 0
	\end{array} \right] 
\end{eqnarray} 

where $0$ and $I$ are $({n\over2}\times{n\over2})$ null matrix and identity
matrix, respectively [11]. It can be shown that if ${\bf D}$ is an
eigenvector corresponding to eigenvalue $\lambda$, then the eigenvector
corresponding to the eigenvalue $-\lambda$ is $S{\bf D}$ [1]. This symmetry
is very evident in our numerical values of the eigenvectors in the case of
coupled quartic oscillators, but is satisfied only approximately in the case
of the highly nonlinear anisotropic Kepler problem. It is to be noted that
the eigenvectors are dependent upon the initial conditions and are only
'local' properties.

\section{Conclusions}

In a recently proposed new method [5] the Lyapunov exponents, are computed
directly, so to say, by utilizing representations of orthogonal matrices,
applied to the tangent map. In this paper, we have established the
connection between this method and a 'differential formulation' of the
standard procedure to compute the Lyapunov spectra.  We have also used the
standard decomposition $SO(4) \sim SO(3) \times SO(3)$ to simplify the
calculations for $n=4$, which are otherwise very involved. It has been
claimed that the new method has several advantages over the existing methods
as it does not require renormalization or reorthogonalization and requires
lesser number of equations. This led us to make a detailed comparison of the
new method with the standard method as well as its differential version , as
regards accuracy and efficiency, by computing the full Lyapunov spectra of
some typical nonlinear systems with 2, 3 and 4 variables. There is
reasonable agreement among the three procedures as far as the values of the
Lyapunov exponents are concerned. However, the standard method seems to
score over the other two, as far as efficiency (as indicated by the CPU time
for a process) is concerned, especially in certain strongly chaotic
situations, and is the most 'robust' procedure. The differential version of
the standard method relies on a stability parameter and seems to demand a
prior estimate of the Lyapunov spectrum. The equations for tangent flow are
nonlinear in this version and highly so in the new method. This is what
makes them less efficient, though the number of coupled differential
equations to be solved is smaller in the new method. However they are still
useful as alternative algorithms for the computation of Lyapunov spectra. We
have also made a comparative study of the computation of the Lyapunov
eigenvectors using the standard method and its differential version. The
eigenvectors converge fairly rapidly (compared to the exponents) and the two
procedures yield essentially identical results.

\vspace{.25in}

{\bf Acknowledgments:}

One of the authors(KR) thanks the Council for Scientific and Industrial
Research, India for financial support through a research fellowship. 

\vspace{.25in}

{\bf References}

\begin{enumerate}
\item See for instance A.J.Lichtenberg and M.A.Lieberman, Regular and
Chaotic dynamics, Springer Velag, New York, 1983.

\item I.Shimada and T.Nagashima, Prog. Theor. Phys. {\bf 61} (1979)
 1605; G.Benettin, L.Galgani, A.Giorgilli and J.M.Strelcyn, Meccanica {\bf
 15} (1980) 9; A.Wolf, J.B.Swift, H.L.Swinney and J.A.Vastano, Physica
{\bf 16D} (1985) 285.

\item I.Goldhirsch, P.L.Sulem and S.A.Orszag, Physica {\bf 27D} (1987) 311.

\item F.Christiansen and H.H.Rugh, Nonlinearity {\bf 10} (1997) 1063.

\item G.Rangarajan, S.Habib and R.D.Ryne, Phys. Rev. Lett.{\bf 80} (1998)
 3747.
\item See for instance, K.Geist, U.Parlitz and W.Lauterborn, Progr. Theor.
Phys. {\bf 83} (1990) 875 and references therein.

\item See for instance, B.G.Wybourne, Classical groups for Physicists, John
and Wiley \& Sons (1974).

\item T.S.Biro, S.G.Matinyan and B.Muller, Chaos and gauge field theory,
World Scientific, Singapore (1994).

\item M.Lakshmanan and R.Sahadevan, Physics Reports {\bf 224} (1993) 1.
 
\item See for instance, M.C.Gutzwiller, Chaos in Classical and Quantum
Mechanics, Springer Verlag, New York (1990).

\item M.Kuwata, A.Harada and H.Hasegawa, J.Phys.A. {\bf 23} (1990) 3227.

\item O.E.R\"ossler, Phys. Lett. {\bf 71A} (1998) 155.  

\item H.Yoshida, Phys Lett {\bf 150A} (1990) 262.

\item See for instance, H.Goldstein, Classical Mechanics, Second Edition,
Addison-Wesley Publishing Co. USA (1980).

\end{enumerate}

\newpage

{\bf Figure Captions}

\vspace{.25in}

Fig.1: Plot of the Lyapunov exponent for the for the coupled quartic
oscillator system. The thin line corresponds to the standard method. Of the
two thick line one with '+' mark corresponds to the differential method and
the one with the 'x' mark corresponds to the new method.

\newpage

\begin{tabular}{|c|ccc|ccc|} \hline

		 & \multicolumn{6}{c|}{Lyapunov spectrum obtained by Standard(1), }\\
System with      & \multicolumn{6}{c|}{ Differential(2) and New(3) methods. Sum of the exponents } \\ 
initial condition& \multicolumn{6}{c|}{ and CPU time in $sec$ are given in  ( ) and [ ] resply.} \\    
                 & \multicolumn{3}{c|}{$t$=10000} &\multicolumn{3}{c|}{$t$=100000} \\		
		 &  1  	&  2 	& 3  	&  1  	&  2 	& 3   \\ \hline

Driven	van der Pol &  0.0985  &  0.0980 &  0.0989  &  0.0987  &  0.0991  &  0.0981   \\
oscillator ($n=2$)  & -6.8494  & -6.8300 & -6.8379  & -6.8411  & -6.8359  & -6.8400   \\	
$z_{1}$=-1.0	    & (-6.7509)&(-6.7321)& (-6.7390)& (-6.7424)& (-6.7368)& (-6.7419) \\
$z_{2}$= 1.0	    &          &         &          & [ 825.56]& [2224.31]& [ 519.22] \\ \hline

Lorenz		   &  0.9022 &  0.9040 &  0.9038 &  0.9051 &  0.9056 & 0.9056 \\
system ($n=3$)	   &  0.0003 &  0.0003 &  0.0001 &  0.0000 &  0.0000 & 0.0000\\
$z_{1}$=0.0        &-14.5691 &-14.5710 &-14.5705 &-14.5718 &-14.5723 &-14.5723\\
$z_{2}$=1.0	   &(-13.667)&(-13.667)&(-13.667)&(-13.667)&(-13.667)&(-13.667)\\
$z_{3}$=0.0        &         &         &         & [1668.7]&[15492.7]& [2394.30]\\ \hline

Anisotropic Kepler&  0.1386 &  0.1343 &  0.1434 &  0.1332 &    &  0.1360 \\
Problem	 ($n=4$)  &  0.0834 &  0.0830 &  0.0860 &  0.0832 &    &  0.0831 \\
$z_{1}$=1.0       & -0.0845 & -0.0817 & -0.0864 & -0.0833 &    & -0.0833 \\
$z_{2}$=2.0       & -0.1375 & -0.1355 & -0.1429 & -0.1331 &    & -0.1357 \\
$z_{3}$=1.0       & (0.0000)& (0.0000)& (0.0000)& (0.0000)&    & (0.0000)\\
$z_{4}$=0.5       &         & [303.25]&         & [201.04]&    & [350.18]\\ \hline

R\"ossler         &  0.1108 &  0.1080 &  0.1125 &  0.1121 &  &  0.1128  \\        
hyperchaos($n=4$) &  0.0224 &  0.0218 &  0.0225 &  0.0196 &  &  0.0214  \\
$z_{1}$=-20.0     & -0.0003 & -0.0003 & -0.0003 & -0.0000 &  & -0.0000  \\
$z_{2}$= 0.0      & -25.9113& -23.7753& -23.9904& -25.1886&  & -24.7527 \\
$z_{3}$= 0.0      &(-25.778)&(-23.646)&(-23.862)&(-25.057)&  &(-24.619) \\
$z_{4}$= 15.0     &         & [4792.61]&        &[5595.99]&  & [1527.68]\\ \hline

Coupled quartic oscr& 0.0009 & 0.0010 & 0.0010 & 0.0001 & 0.0001 & 0.0001 \\
($n=4,\alpha=6$)    & 0.0008 & 0.0008 & 0.0008 & 0.0001 & 0.0001 & 0.0001 \\
$z_{1}$= 0.8        & -0.0008&-0.0008 &-0.0008 &-0.0001 &-0.0001 &-0.0001 \\
$z_{2}$= 0.5	    & -0.0009&-0.0010 &-0.0010 &-0.0001 &-0.0001 &-0.0001 \\
$z_{3}$= 1.0	    &(0.0000)&(0.0000)&(0.0000)&(0.0000)&(0.0000)&(0.0000)\\
$z_{4}$= 1.3	    &        &        &        &[492.09]&[5453.3]&[803.49]\\ \hline

Coupled quartic oscr& 0.1892 & 0.2096 & 0.1739 & 0.1738 & 0.1793 & 0.1806\\
($n=4,\alpha=8$)    & 0.0011 & 0.0011 & 0.0010 & 0.0001 & 0.0001 & 0.0001\\
$z_{1}$= 0.8        &-0.0011 &-0.0011 &-0.0013 &-0.0001 &-0.0001 &-0.0001\\
$z_{2}$= 0.5	    &-0.1892 &-0.2095 &-0.1795 &-0.1738 &-0.1793 &-0.1806\\
$z_{3}$= 1.0	    &(0.0000)&(0.0000)&(0.0000)&(0.0000)&(0.0000)&(0.0000)\\
$z_{4}$= 1.3	    &        &        &        &[492.09]&[7658.6]&[39012.8]\\ \hline

\end{tabular}

\vspace{.25in} 

{\bf Table 1:} Comparison of the Lyapunov spectrum and the computational
time required to evaluate them with three different methods for some of the
systems with $n=2, 3$ and 4.

\vspc

\begin{tabular}{|c|c|c|} \hline

System with	 	& \multicolumn{2}{c|}{Lyapunov eigenvectors}     \\ 
initial condition    	&  Standard method  &  Differential method  \\ \hline
Driven	van der Pol	&		&   	\\
oscillator ($n=2$)	& $D_{1}$( 0.894, 0.447)  & $D_{1}$( 0.894, 0.447)  \\
$z_{1}$=-1.0   		& $D_{2}$(-0.447, 0.894)  & $D_{2}$(-0.447, 0.894)  \\
$z_{2}$= 1.0   	     	&       	  &	         \\	\hline
Lorenz  	&   			& \\	
system ($n=3$)  & $D_{1}$( 0.004, 0.040,-0.999) & $D_{1}$( 0.004, 0.040,-0.999) \\
$z_{1}$=0.0  	& $D_{2}$(-0.789,-0.614,-0.028) & $D_{2}$(-0.789,-0.614,-0.028)  \\
$z_{2}$=1.0     & $D_{3}$(-0.614, 0.788, 0.029) & $D_{3}$(-0.614, 0.788, 0.029) \\ 
$z_{3}$=0.0	&	     		&    \\	\hline
Anisotropic Kepler&    					& \\
Problem	 ($n=4$)  & $D_{1}$( 0.230, 0.139,-0.868,-0.417)& $D_{1}$( 0.233, 0.136,-0.863,-0.428)\\
$z_{1}$=1.0       & $D_{2}$(-0.291, 0.262,-0.427, 0.815)& $D_{2}$(-0.288, 0.263,-0.438, 0.810)\\
$z_{2}$=2.0	  & $D_{3}$(-0.373, 0.854, 0.187,-0.309)& $D_{3}$(-0.371, 0.855, 0.188,-0.309)\\
$z_{3}$=1.0	  & $D_{4}$( 0.850, 0.427, 0.171, 0.255)& $D_{4}$( 0.851, 0.425, 0.170, 0.256)\\
$z_{4}$=0.5	  &		&    \\	\hline
R\"ossler	  &    					& \\
hyperchaos ($n=4$)& $D_{1}$( 0.660, 0.081,-0.051, 0.745)& $D_{1}$( 0.660, 0.081,-0.052, 0.745)\\
$z_{1}$=-20.0     & $D_{2}$(-0.749, 0.115, 0.022, 0.653)& $D_{2}$(-0.749, 0.111, 0.022, 0.653)\\
$z_{2}$= 0.0 	  & $D_{3}$(-0.014,-0.928, 0.347, 0.137)& $D_{3}$( 0.030, 0.991, 0.005,-0.134)\\
$z_{3}$= 0.0 	  & $D_{4}$(-0.058,-0.345,-0.936, 0.025)& $D_{4}$( 0.050,-0.003, 0.998, 0.025)\\
$z_{4}$= 15.0	  &		&    \\	\hline
Coupled quartic oscr.  &    					& \\
($n=4,\alpha=6$)  & $D_{1}$( 0.687, 0.685, 0.162, 0.182)& $D_{1}$( 0.684, 0.684, 0.178, 0.185)\\
$z_{1}$= 0.8      & $D_{2}$(-0.223, 0.241, 0.672,-0.663)& $D_{2}$(-0.234, 0.241, 0.666,-0.666)\\
$z_{2}$= 0.5	  & $D_{3}$( 0.670,-0.666, 0.231,-0.232)& $D_{3}$( 0.666,-0.666, 0.241,-0.234)\\
$z_{3}$= 1.0	  & $D_{4}$(-0.170,-0.173, 0.684, 0.688)& $D_{4}$(-0.185,-0.178, 0.684, 0.684)\\
$z_{4}$= 1.3	  &		&    \\	\hline
Coupled quartic oscr.  &    					& \\
($n=4,\alpha=8$)  & $D_{1}$( 0.503,-0.351, 0.581,-0.535)& $D_{1}$( 0.503,-0.352, 0.583,-0.533)\\
$z_{1}$= 0.8      & $D_{2}$( 0.634, 0.744, 0.080, 0.196)& $D_{2}$( 0.635, 0.740, 0.085, 0.204)\\
$z_{2}$= 0.5	  & $D_{3}$(-0.084,-0.192, 0.638, 0.741)& $D_{3}$(-0.088,-0.201, 0.633, 0.743)\\
$z_{3}$= 1.0	  & $D_{4}$( 0.581,-0.536,-0.498, 0.356)& $D_{4}$( 0.580,-0.536,-0.502, 0.351)\\
$z_{4}$= 1.3	  &		&    \\	\hline

\end{tabular} 

\vspc

{\bf Table 2:} Comparison of the Lyapunov eigenvectors computed using the
differential and the standard methods for some systems with $n=2,3$ and 4.
The eigenvectors are at $t=1000$ for the differential method and at $t=
1000, 35, 150, 170, 200, 20 $ respectively form top to bottom for the
standard method.

\end{document}